**CARES: A Conversational AI System for Regulation-Grounded Safety Reporting in Construction Education**

Fan Yang[1]; Jiabin Wu[2]; Yuan Tian[3]; and Jiansong Zhang, Ph.D., A.M.ASCE[4]

[1] School of Construction Management Technology, Purdue Univ., West Lafayette, IN, USA. Email: yang2352@purdue.edu
[2] Chair of Computing in Civil and Building Engineering & TUM Georg Nemetschek Institute, Technical University of Munich, 80333, Munich, Germany. Email: j.wu@tum.de
[3] Department of Computer Science, Purdue Univ., West Lafayette, IN, USA. Email: tian211@purdue.edu
[4] School of Construction Management Technology, Purdue Univ., West Lafayette, IN, USA. Email: zhan3062@purdue.edu

## ABSTRACT

Construction safety reporting often relies on manual logs and static templates that provide limited feedback and leave daily activities disconnected from relevant regulations. This paper introduces CARES (Conversational AI Reporting for Enhanced Safety), a conversational AI system that integrates regulatory guidance into daily reporting to support construction safety education. CARES combines proactive multi-agent dialogue, retrieval-augmented generation (RAG), and automated report generation. The system guides users through reporting tasks, retrieves relevant regulatory passages using hybrid retrieval, and converts conversations into structured daily reports. Regulatory sources and the evolving report are displayed alongside the dialogue to support user review and correction. A preliminary evaluation involving 15 construction management students assessed retrieval quality, response faithfulness, and conversational relevance. CARES achieved an overall faithfulness score of 0.74 and an answer relevance score of 1.00, while initial retrieval ranking remained an area for improvement. These results provide preliminary evidence of the technical feasibility of regulation-grounded conversational reporting. The study highlights opportunities to integrate regulatory knowledge into routine documentation, with future work needed to evaluate effects on report quality, safety awareness, and learning outcomes.

## INTRODUCTION

Within construction safety education context, numerous safety events stem not only from technical issues but also from human factors including problematic communication, misunderstanding of regulations and inadequate knowledge. In addition to traditional training and classroom learning, construction safety should also be reinforced within the regular workflows whereby site activities are documented, reported, and assessed. The most typical example is daily construction reporting. Activities on-site, equipment and material use, labor deployment, observation of progress and safety-related incidents are recorded in daily reports. Such reports are needed to coordinate and keep records of a project, though they are usually prepared manually once the work has already taken place. Consequently, reporting is often a retrospective and repetitive documentation process as opposed to an active system in strengthening safety knowledge in the workplace. Promoting reflection on daily tasks and reinforcement on the safety regulations could facilitate a more positive safety awareness and professional responsibility.

A major difficulty is that construction safety regulations are not readily available and accessible. The regulatory resources are usually found in long, text-filled books or on the Internet which are not interconnected with the everyday site work. This knowledge is complicated and difficult to interpret, thus making it difficult to retrieve when workers require it. In this sense, daily reporting is a good opportunity that has not been fully utilized. Because workers already need to write about their work, equipment and safety observations, the reporting process is a logical means to tie up the day-to-day work with the applicable safety regulations. Present day record keeping is nearly a passive form of reporting. The gap in the research is also evident: not many systems are capable of generating reports based on conversational AI and automatically linking activity descriptions to relevant regulations.

Recent developments in the field of Artificial Intelligence (AI) provide new opportunities to improve safety education by using Large Language Model (LLM)-driven conversational systems. Such systems have been shown to support learning and to engage users in a natural language dialogue, as well as to provide explanations in a variety of fields. And Retrieval-Augmented Generation (RAG) opens a promising direction to integrate domain knowledge into conversational AI systems. Conversational assistance can enable users to note on-site operations via natural-language dialogue as opposed to creating manual reports, and RAG can map such descriptions to structured regulatory material and offer real-time context-based feedback.

To overcome these shortcomings, we proposed a conversational reporting system that combines report generation based on conversation and regulation retrieval. By doing so, the reporting will not only give the picture of what occurred on the ground but also connect site users to pertinent regulatory information as users narrate their work. The proposed system will identify activities from structured conversations led by a proactive AI agent and automatically format them into standardized digital reports. In addition, connecting the records of conversations and the related regulatory references, the system will turn the traditional documentation into a type of real-time knowledge exchange.

By so doing, this paper reframes reporting from a passive record keeping activity to an active, human-oriented process of safety support. This process can be integrated into daily construction processes, and it allows users to have consistent feedback and regulatory reinforcement in their daily reporting practices.

## BACKGROUND

### On-Site Construction Safety and Regulatory Compliance

Construction projects are governed by a wide range of safety regulations, including building codes, Occupational Safety and Health Administration (OSHA) standards, and Environmental Protection Agency (EPA) requirements. These regulatory documents contain numerous provisions on equipment use, work procedures, and protective measures for different site conditions (Zhang 2017; Wang 2023). Following these requirements is essential for ensuring on-site safety and reducing fatalities during construction.

However, maintaining consistent regulatory compliance on construction sites remains challenging (Wang 2021). Unsafe worker behavior is a major cause of construction accidents (Li 2026), making effective safety training an important means of improving on-site safety (Liao 2025). Research on virtual reality training further suggests that training format can influence learners' perceived utility and affective reactions (Yang et al. 2026). Such training depends on providing workers with clear and relevant safety information that can guide their actions in specific work situations. In practice, however, this information is often scattered across different stakeholders, tools, and project phases, creating information silos that make safety knowledge difficult to access, share, and apply (Li 2022). As a result, the availability of safety regulations alone does not ensure their effective application on site. Workers and safety personnel must still identify relevant rules, interpret their requirements, and translate them into appropriate actions. Current reporting and training practices provide limited support for this process, leaving a gap between written requirements and on-site implementation.

**Construction Safety Reporting and Regulatory Awareness**

Construction reporting is an essential part of project management, as well as the safety governance in the Architecture, Engineering, and Construction (AEC) sector. Site activities, progress, resource use, and observations of safety-related aspects are regularly documented using daily reports, field logs, and inspection records. The purposes of these records are to coordinate a project, document contracts, and comply with regulations, as well as to analyze the incident after it has occurred. Previous studies have highlighted the significance of proper and prompt reporting in enhancing transparency and accountability of a project (Eastman 2011; Love et al. 2016).

Although it is important, construction reporting is still very manual and retroactive. Reports are usually done by field personnel at the end of the workday and are often based on memory or on some fragmented notes. The risks that are brought about by this process include incomplete, inconsistent, and inaccurate documentation. In construction domain, research studies have revealed that manual reporting systems tend to lose information and become inconsistent and unreliable among different people, especially in dynamic and time-sensitive settings (Golparvar-Fard et al. 2015; Kim et al. 2013). Consequently, reporting is often approached as a managerial procedure, as opposed to a dynamic tool for promoting safety awareness and reinforcing knowledge in the course of work.

Concurrently, regulatory frameworks like those that have been developed by OSHA are very essential in construction safety management. These regulations stipulate safety standards, mandatory procedures, and compliance requirements. Regulatory knowledge is, however, usually contained in vast textual documents that are not readily available when it comes to daily operations. According to previous studies, employees and supervisors tend to struggle with the interpretation and implementation of regulatory items in a specific situation, especially when they are experiencing time pressure and cognitive load conditions (Lingard et al. 2015). This lack of connection between regulatory knowledge and on-site activity restrains the success of safety training and compliance activities.

### Conversational AI with Retrieval-Augmented Generation

Conversational AI has been studied in the AEC industry as a natural language interface to project information. A review by Saka et al. (2023) found that existing conversational systems for AEC remain limited in deployment and focus mainly on information retrieval from Building Information Modeling (BIM)-based design data. For example, Zheng and Fischer (2023) developed a prompt-based virtual assistant that interprets natural language queries and retrieves information from BIM databases using LLMs. Yang et al. (2025) developed a multi-agent conversational system that integrates domain knowledge through RAG to provide construction workers with safety guidance and emotional support. These applications demonstrate the potential of conversational AI for information access and worker support, while the integration of regulatory guidance with structured daily reporting remains less explored.

Interactive natural language interfaces have grounded AI-generated queries in editable explanations and visual feedback to support user understanding and correction (Tian et al. 2023; Tian et al. 2024), offering design insights for grounded conversational reporting.
The RAG-based approaches ground LLM outputs in passages retrieved from an external knowledge base, improving factual accuracy and providing sources for generated claims (Lewis et al. 2020). Recent studies have applied RAG to construction safety knowledge. Uhm et al. (2025) showed that retrieval grounding improves the accuracy of generated safety information while mitigating hallucination. Lee et al. (2024) found that both RAG and fine-tuned LLMs outperform a general-purpose GPT-4 baseline in safety knowledge retrieval. Baek et al. (2025) demonstrated that a RAG-enhanced LLM can generate safety risk guidance comparable in quality to that of experienced practitioners. As RAG systems mature, automated methods have also emerged to evaluate their output quality. The RAGAS framework (Es et al. 2024) measures whether generated responses are supported by the retrieved context, and whether responses address the user input.

However, integrating regulation-grounded assistance into routine daily reporting remains underexplored. To bridge the gap, CARES leverages daily reporting, where workers typically describe their on-site activities, and integrates regulation retrieval directly into the reporting dialogue. The agent proactively elicits activity descriptions, presents relevant regulations with sources, and automatically transforms the dialogue into a structured daily report.

## METHODOLOGY

We propose a conversational AI assistant, which assists on-site construction workers and trainees to raise regulatory awareness and generate compliant safety documentation through interactive, knowledge-based dialogue. It is structured around three coordinated mechanisms: (1) multi-agent conversational coordination, (2) retrieval-augmented generation (RAG) over regulatory knowledge and (3) integrated digital documentation and reporting. As the base LLM across all agents, we use gpt-4o (OpenAI 2024). The user interface of the proposed CARES system is shown in Figure 1.

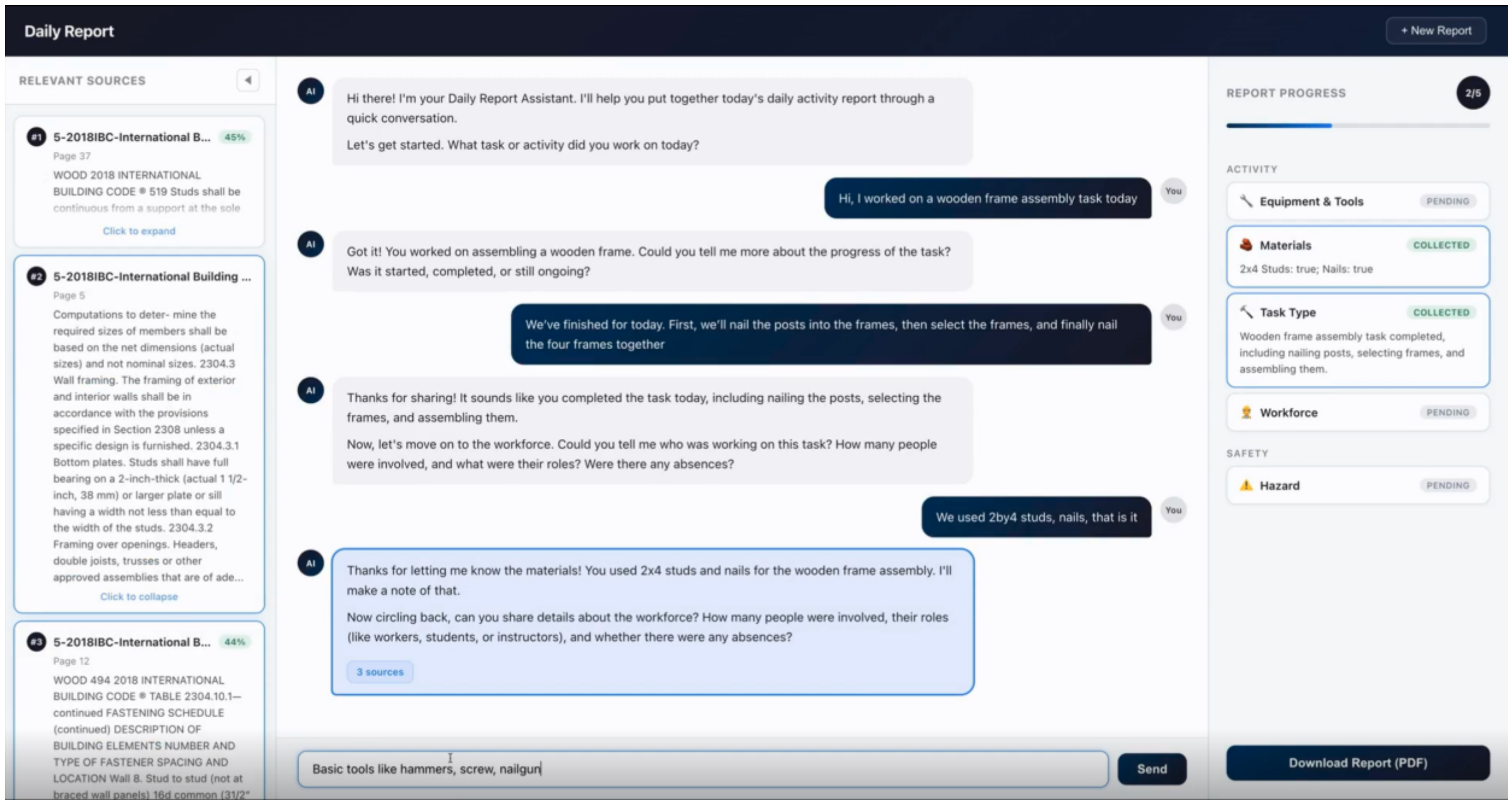

**Figure 1. User interface of the CARES system.**

**Multi-Agent Conversational Coordination.** Rather than relying on a single monolithic prompt, we break down the assistant into specialized agents, which share the same session state and coordinate with each other via structured message passing.

There is a dialogue agent, which communicates with the user, requesting one topic at a time, taking notes of previous input, and guiding the conversation toward the next unfilled report field.

A knowledge agent processes RAG queries against the regulatory corpus, returning ranked passages with source attributions.

A structured JSON snapshot of the report is re-derived by an extraction agent that re-reads the entire conversation after every turn and re-derives a structured JSON snapshot of the report, in the spirit of recent prompt-based zero-shot information extraction (Wei et al. 2023). Re-extracting in a fresh fashion (as opposed to incrementally) makes the snapshot resilient to user corrections later in the discussion and avoids the accumulation of extraction errors across turns.

Lastly, the snapshot is serialised into a standardised PDF, on request, by a so-called reporting agent. The agents communicate over a shared session object keyed by Universally Unique Identifier (UUID), and all remote model calls are wrapped with exponential-backoff retries so that temporary API failures do not break the conversational loop. This breakdown allows the assistant to be conversational and responsive to in-situ context, rather than acting off of a fixed, rule-based script.

**RAG over Regulatory Knowledge.** A knowledge agent transforms a fixed corpus of construction-safety regulations into a dynamically accessible body of knowledge. The existing corpus is made up of authoritative documents gathered by international and national authorities, including OSHA fact sheets on residential construction (Occupational Safety and Health Administration 2012) and the Wood Frame Construction Manual (American Wood Council 2012). On startup, the corpus is chunked and embedded with the *text-embedding-3-large* model, and the resulting vector index is stored to disk to avoid re-embedding across launches.

A two-stage retrieval pipeline is adopted at query time. The first stage uses hybrid retrieval, combining dense vector and BM25 keyword retrieval through reciprocal rank fusion, to select the top eight candidate passages for the user's latest message. The second step uses a re-ranking of the candidates using backbone LLM-as-judge, where the backbone LLM by itself classifies each candidate as relevant or not. Up to four "yes" chunks (in the fused ranking order) are concatenated into the prompt of the dialogue agent as a block of context. Most importantly, the metadata of each retained chunk (source file, page label, similarity score, and snippet) are propagated to the front-end and rendered as expandable source cards, providing auditable, page-level provenance between every regulatory claim and the original document, and transforming otherwise opaque, text-heavy regulations into a transparent and conversation-ready resource.

**Digital Documentation and Reporting.** The dialogue itself is formalized by the third mechanism.

The target report adheres to a fixed schema with two sections, denoted by the dialogue policy and the rendered output, respectively. The extraction agent, as the conversation continues, continuously identifies activity- and compliance-relevant statements made by the user and writes them into the structured snapshot which the front-end renders in real time by a set of field cards along with a completion-ratio progress bar so that the user can see precisely what fields have been filled.

When the user requests a download, the reporting agent serializes the snapshot into a standardized PDF which has been prepended with a brief two-to-three-sentence narrative summary of the work done on that day generated by another call to the backbone LLM over the conversation. The front-end displays a 3-panel interface where the chat, the live report panel, and the regulatory source panel are visible side-by-side, such that the process of writing out the daily report also provides a situation in which situated learning can occur regarding the regulations that underlie each safety question.

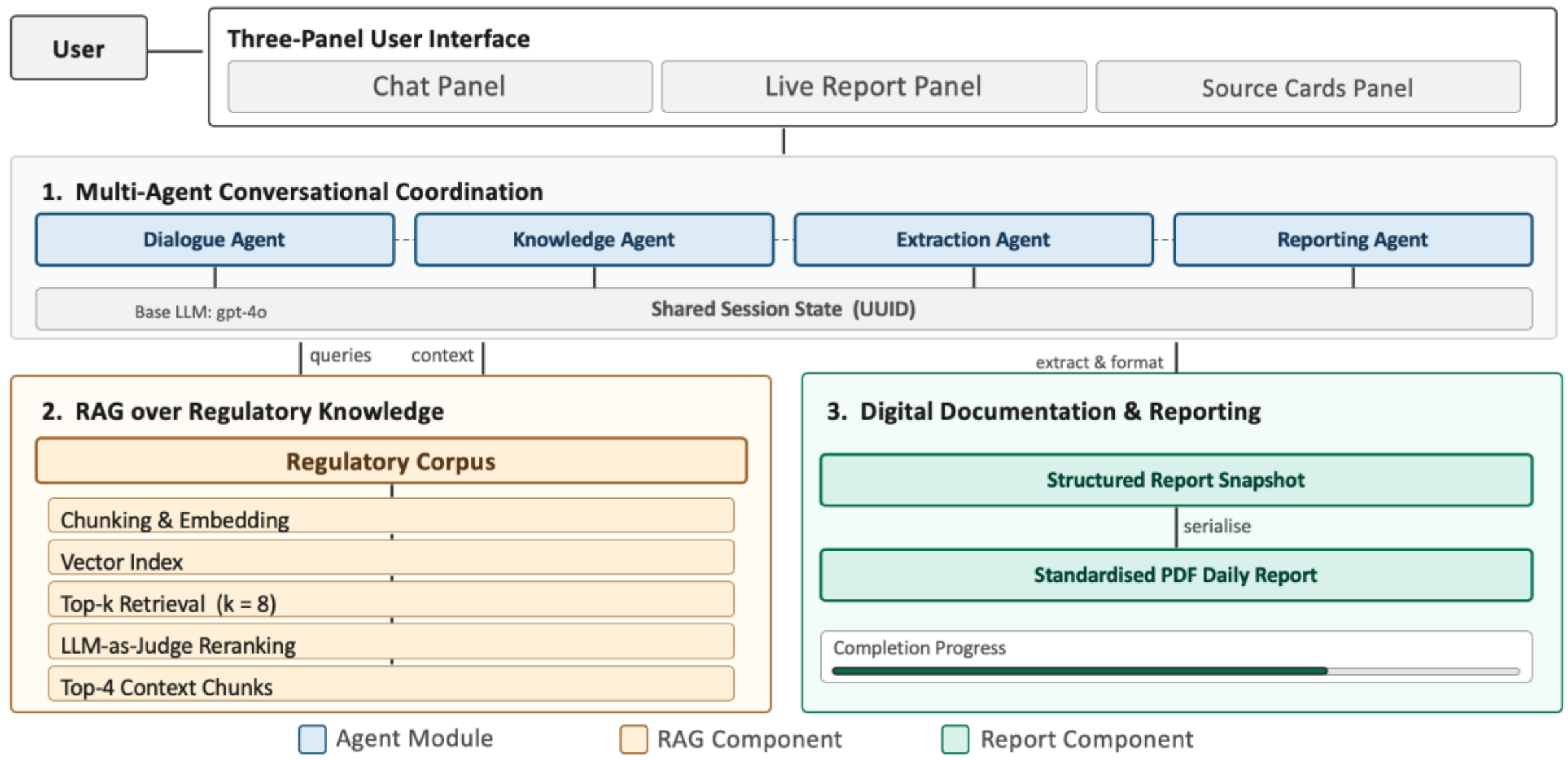

**Figure 2. Overall architecture and workflow of the CARES system, integrating the three-panel user interface, multi-agent conversational coordination, regulation-grounded RAG, and automated digital report generation**

Figure 2 demonstrates the overall architecture of the system. The system bridges the gap between informal everyday discourse and formal compliance records and incrementally accumulates a project-level base of knowledge of safety practice, which can be reused.

**Measurement** The technical quality of the RAG system is assessed using three complementary automated measures: a retrieval precision proxy, faithfulness, and answer relevance. Together, these measure whether (1) relevant material appears near the top of the retrieval ranking, (2) whether generated content is supported by retrieved evidence, and (3) whether the assistant appropriately responds to the participant's immediately preceding message.

**Retrieval Precision Proxy@4.** Retrieval Precision Proxy@4 estimates the concentration of relevant information near the top of the retrieval ranking. Dense vector and BM25 keyword retrieval are combined using reciprocal rank fusion, and the top eight fused candidates are classified as relevant or not relevant by the retrieval-stage backbone model (gpt-4o). Up to four positively judged chunks are then provided to the assistant in response context. The proxy score is the proportion of relevant candidates among the first four fused results. Because relevance is determined by the system's retrieval-stage model rather than human annotators, the measure is treated as a proxy rather than a human-labeled Precision@4 benchmark. It is not calculated when no candidates are retrieved.

**Faithfulness.** Faithfulness measures the extent to which factual claims in a generated response are supported by the retrieved context. For each eligible turn, gpt-4o-mini extracts factual claims while excluding conversational content such as acknowledgements, questions, and meta-commentary. The claims are then independently evaluated against the retrieved passages by three LLM judges: gpt-4o-mini, Gemini 2.5 Flash, and Claude 3 Haiku. For each judge, turn-level faithfulness is the proportion of extracted claims judged as supported, yielding a score from 0 to 1. Session-level and pooled scores are obtained by averaging eligible turn-level scores, with majority-vote summaries also reported across the three judges. Faithfulness is not calculated when no context is retrieved, or no factual claims are extracted.

**Answer Relevance**. Answer relevance evaluates whether a response appropriately addresses the user's immediately preceding message. A response is considered relevant if it directly answers the message, accurately acknowledges its content, or asks a related follow-up question, including questions used to complete missing report fields. Each response is independently classified as relevant or not relevant by the same three LLM judges. For each judge, the aggregate score is the proportion of eligible turns classified as relevant, with majority-vote summaries also reported across the three judges.

**User Study Procedures.** The user study was designed to evaluate the preliminary technical feasibility of CARES. As illustrated in Figure 3, the procedure consisted of four stages: participant recruitment and orientation, AI-enabled conversation, automated report generation, and data collection and system evaluation.

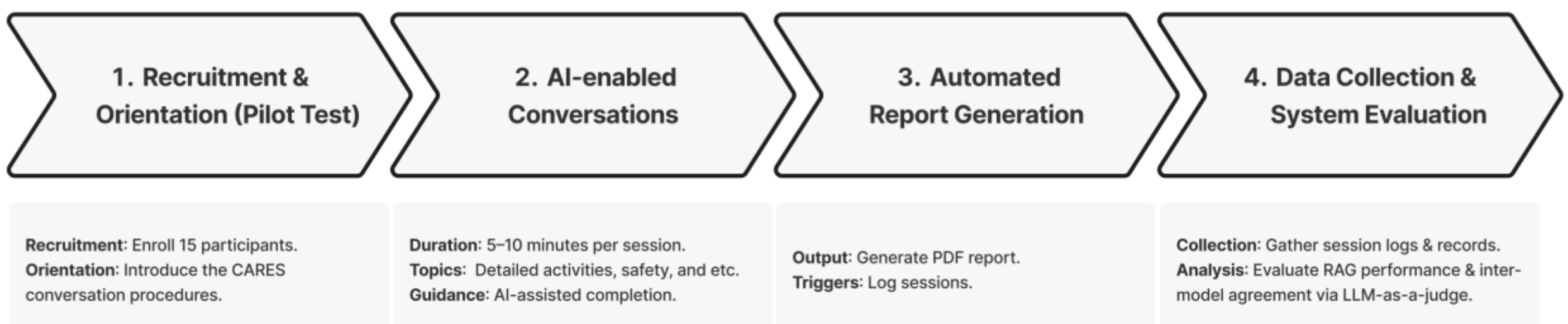

**Figure 3. User study procedure and technical evaluation workflow.**

**Step 1:** With the purposes of system assessment, 15 participants were recruited. They are instructed to interact with CARES based on their recent construction experience. Each of them completed one CARES reporting session and generated one daily report, results in 15 reports in total.

**Step 2:** Each participant received a brief orientation to the CARES interface and interacted with the system for approximately 5–10 minutes. They described the construction activities performed, safety measures observed or applied, and hazards encountered, while the dialogue agent guided them toward completion of all report fields.

**Step 3:** After completing the interaction, participants downloaded the auto-generated PDF report, which also triggered storage of the conversational session and its evaluation data. This evaluation aimed only to establish the prototype's end-to-end technical feasibility, rather than comparative usability or learning outcomes, which will be addressed in a larger user study in the future.

**Step 4:** During data collection, participants interacted with CARES in real time, while the retrieved candidate pool, relevance judgments, and accepted context were recorded with each conversation and exported at session completion. Faithfulness and Answer Relevance were then computed offline from the saved sessions, preventing judge-model latency or failures from affecting the participant's experience and allowing the evaluation to be rerun when needed. Metrics were aggregated at the turn, session, and pooled-study levels, while evaluation failures were logged separately from negative judgments to avoid conflating technical issues with poor system performance.

## RESULTS AND ANALYSIS

The preliminary evaluation examined the technical performance of CARES across 15 reports. The analysis focused on the retrieval and conversational performance of the RAG-assisted interaction using the three measures defined in the Methodology section. Table 1 summarizes the pooled evaluation results, while Figure 4 presents an example of a structured daily report generated from a CARES conversation.

**Table 1. Technical evaluation of the RAG component using LLM-as-a-judge metrics.**

| Metric | GPT-4o-mini Judge | Gemini 2.5 Flash Judge | Claude 3 Haiku Judge | Overall Score |
|---|---|---|---|---|
| Retrieval Precision Proxy@4 | — | — | — | **0.17** |
| Faithfulness | 0.77 | 0.31 | 0.77 | **0.74*** |
| Answer Relevance | 0.99 | 0.99 | 1.00 | **1.00*** |

*The overall scores for Faithfulness and Answer Relevance are based on majority voting across the three LLM judges.

The pooled Retrieval Precision Proxy@4 was 0.17, indicating that relevant regulatory information was not consistently concentrated within the first four positions of the fused retrieval ranking. This result suggests that the initial ranking remains an area for further improvement. However, the proxy evaluates the first four fused results before the subsequent relevance filtering stage. In the CARES pipeline, the top eight retrieval candidates are further evaluated, and up to four positively judged passages are supplied to the dialogue agent. Therefore, the proxy primarily reflects the quality of the initial retrieval ranking rather than the quality of the final context used for response generation.

For Faithfulness, the overall majority-vote score was 0.74, indicating that most factual claims evaluated during eligible conversational turns were supported by the retrieved regulatory context. GPT-4o-mini and Claude 3 Haiku produced identical scores of 0.77, whereas Gemini 2.5 Flash produced a lower score of 0.31. The variation across judges indicates that automated faithfulness evaluation is sensitive to the evaluator model. Nevertheless, the agreement between GPT-4o-mini and Claude 3 Haiku, together with the majority-vote result, provides preliminary evidence that CARES generally produced responses grounded in the retrieved regulatory material.

Answer Relevance showed the strongest and most consistent performance. GPT-4o-mini and Gemini 2.5 Flash each produced a score of 0.99, while Claude 3 Haiku produced a score of 1.00, resulting in an overall majority-vote score of 1.00. The consistency across the three judges indicates that CARES reliably responded to the participant's immediately preceding message. Since contextually appropriate follow-up questions were also considered relevant under this metric, the result further suggests that the dialogue agent was able to maintain conversational continuity while guiding users toward completion of the reporting task.

Taken together, the results show different levels of performance across the retrieval and generation stages. The comparatively low Retrieval Precision Proxy@4 suggests room for improvement in ranking regulatory passages, while the higher Faithfulness score indicates that the downstream filtering and generation process was generally able to produce responses supported by the selected context. At the conversational level, the near-perfect Answer Relevance scores demonstrate that the system consistently maintained alignment with user input. These findings provide initial support for the technical feasibility of combining regulatory retrieval with proactive conversational reporting.

## DISCUSSION

CARES provides a new mechanism for embedding regulatory knowledge into daily reporting, potentially supporting regulatory awareness during construction activities. As users describe their work, the system retrieves applicable regulatory passages and presents them alongside the conversation and evolving report. Reporting can therefore become an opportunity for situated exposure to regulatory knowledge rather than remaining solely a retrospective documentation task.

Another contribution of this work is the use of proactive agentic conversation to structure and guide reporting. CARES is not a conventional chatbot that operates reactively. Instead, it maintains the state of the report, identifies incomplete fields, and asks targeted follow-up questions to elicit missing activity and safety information. The dialogue agent consequently serves not only as a question-answering interface but also as a process coordinator. The extraction agent's repeated reconstruction of the report from the full conversation further allows corrections made later in the interaction to be reflected in the structured report.

Lastly, CARES could contribute to lowering the barriers to regulatory knowledge learning. Construction regulations are commonly distributed across lengthy, technically written documents, requiring users to locate, interpret, and relate individual provisions to a specific work activity. CARES lowers this barrier by converting the regulatory corpus into a conversationally accessible resource. Its two-stage retrieval and reranking process identifies potentially relevant passages, while page-level source cards allow users to inspect the provenance of the information presented. The system therefore reduces the effort required to search multiple documents manually and provides a more transparent alternative to relying exclusively on an LLM's parametric knowledge.

While these preliminary findings are promising, several limitations need to be acknowledged.

First, the evaluation involved only 15 participants. This is the preliminary stage of the project, mainly focused on technical feasibility. The findings need to be examined in a larger-scale user study in the future to demonstrate the system's robustness or effectiveness across diverse users and real-world construction contexts.

Second, the current regulatory corpus covers a limited set of documents and clauses relevant to the selected construction exercise. Expanding the knowledge base to encompass broader OSHA requirements, building codes, manufacturer guidance, project-specific safety plans, and jurisdiction-dependent standards will create additional retrieval challenges.

Finally, the present evaluation relied partly on LLM-based judges, including a retrieval precision proxy based on model-generated relevance labels. The use of multiple independent judges improves the robustness of the faithfulness and answer-relevance assessment, but it does not replace evaluation by construction safety and regulatory experts.

## CONCLUSION

Construction reporting remains largely retrospective and disconnected from the regulatory knowledge needed to interpret on-site activities and safety practices. To address this gap, this study developed CARES, a conversational AI system that integrates proactive multi-agent dialogue, retrieval-augmented access to authoritative construction regulations, and automated generation of structured daily reports. Rather than requiring users to independently search complex regulatory documents or complete static reporting templates, CARES guides users through required reporting fields, connects their activity descriptions with source-attributed regulatory information, and converts the conversation into formal documentation. The preliminary evaluation of 15 reporting sessions demonstrates the end-to-end technical feasibility of CARES, showing reliable conversational relevance and generally grounded responses. Taken together, CARES demonstrates how construction reporting systems can extend beyond information recording to provide proactive guidance, regulation-grounded assistance, and automated generation of structured reports.

The next stage of this research will extend the technical assessment through a large-scale user study, deploying CARES across diverse project environments with a broader cohort of construction professionals and students to compare report quality, completion time, and cognitive workload against conventional workflows. This will enable the research team to scale the regulatory corpus to incorporate project-specific safety mandates, and technical specifications to test retrieval robustness.

## ACKNOWLEDGEMENT

We used AI assistants (e.g., ChatGPT) for limited language editing and phrasing suggestions during the writing process. All technical content, experimental design, and conclusions were conducted by the authors.

## APPENDIX

### Example of a CARES-Generated Daily Report

This appendix provides an example of a structured daily report generated from one of the evaluated CARES conversations. The report was automatically produced from the participant's conversational input, with activity descriptions, safety observations, and other reporting information extracted and organized according to the predefined report structure.

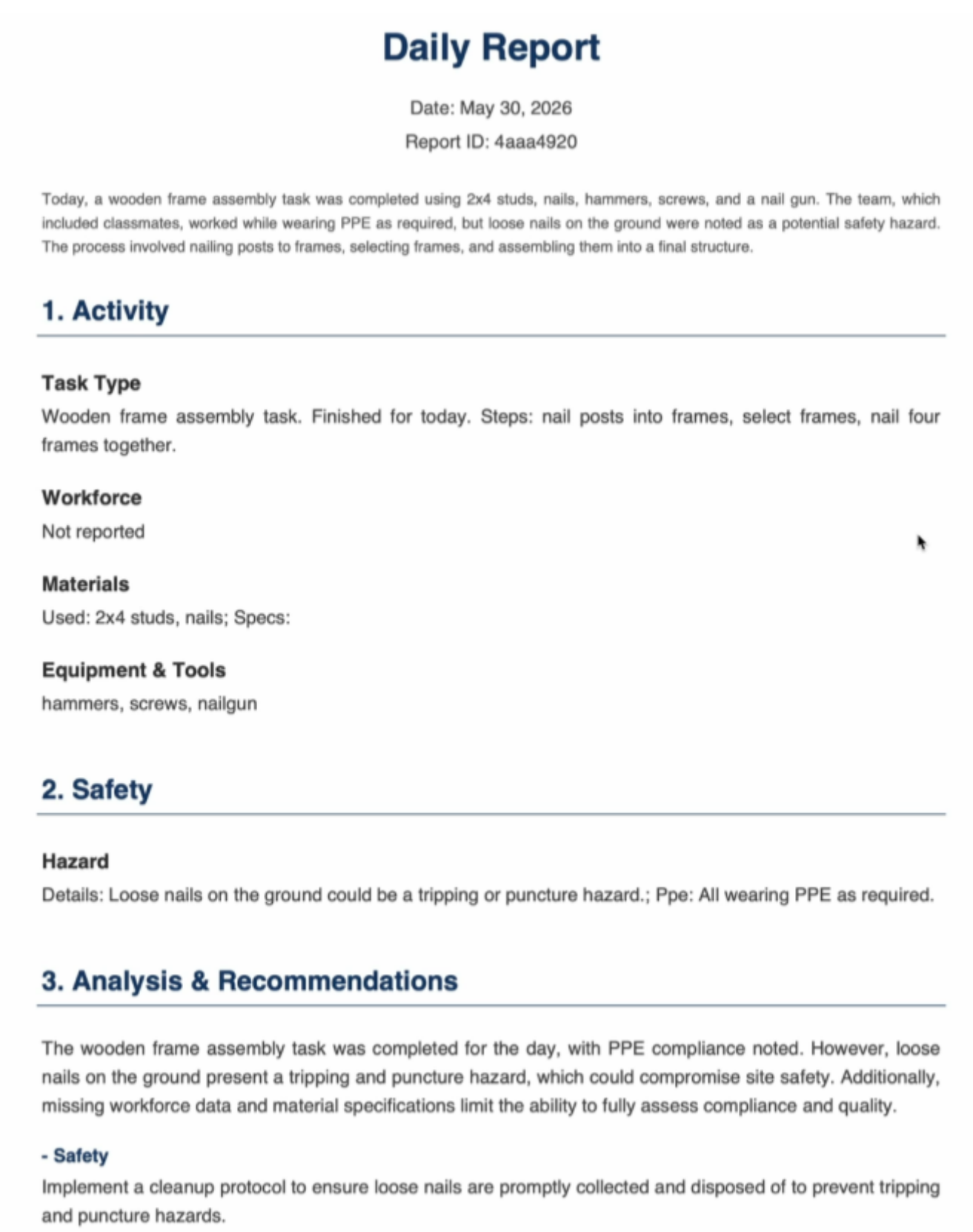

Daily Report

Date: May 30, 2026

Report ID: 4aaa4920

Today, a wooden frame assembly task was completed using 2x4 studs, nails, hammers, screws, and a nail gun. The team, which included classmates, worked while wearing PPE as required, but loose nails on the ground were noted as a potential safety hazard. The process involved nailing posts to frames, selecting frames, and assembling them into a final structure.

1. Activity

Task Type

Wooden frame assembly task. Finished for today. Steps: nail posts into frames, select frames, nail four frames together.

Workforce

Not reported

Materials

Used: 2x4 studs, nails; Specs:

Equipment & Tools

hammers, screws, nailgun

2. Safety

Hazard

Details: Loose nails on the ground could be a tripping or puncture hazard.; Ppe: All wearing PPE as required.

3. Analysis & Recommendations

The wooden frame assembly task was completed for the day, with PPE compliance noted. However, loose nails on the ground present a tripping and puncture hazard, which could compromise site safety. Additionally, missing workforce data and material specifications limit the ability to fully assess compliance and quality.

- Safety

Implement a cleanup protocol to ensure loose nails are promptly collected and disposed of to prevent tripping and puncture hazards.

**Figure 4. Example of a structured daily report generated from a CARES conversation.**